\documentstyle[11pt,epsfig]{article}
\def\xname #1#2#3#4{%    x-axis label, y-axis label, filename, size
  \dimen0=#4\hsize \dimen0=0.9\dimen0%
  \dimen1=0.02\hsize%
  \hbox{\vbox{\large
    \hbox to0pt{\hskip\dimen1\strut #1\hss}
    \hbox{\epsfig{%bbllx=5.4cm,bblly=1.9cm,bburx=28.7cm,bbury=19.6cm,
          file=#3,width=#4\linewidth%,height=0.9\textheight
    }}%
    \hbox to\dimen0{\hss\strut #2}
  }}%
}
\newlength{\dinwidth}
\newlength{\dinmargin}
\def\eqalign#1{
\null \,\vcenter {\openup \jot \ialign {\strut \hfil $\displaystyle {
##}$&$\displaystyle {{}##}$\hfil \crcr #1\crcr }}\,}

\newcounter{subequation}[equation]
\makeatletter
\expandafter\let\expandafter\reset@font\csname reset@font\endcsname
\newenvironment{subeqnarray}
  {\arraycolsep1pt
    \def\@eqnnum\stepcounter##1{\stepcounter{subequation}{\reset@font\rm
      (\theequation\alph{subequation})}}\eqnarray}%
  {\endeqnarray\stepcounter{equation}}
\makeatother

\def\abs#1{\left| #1\right|}
\def\be#1{\begin{equation}\label{#1}}
\def\ee{\end{equation}}
\def\ba#1{\begin{subeqnarray}\label{#1}}
\def\ea{\end{subeqnarray}}

\def\e{{\rm e}}

\begin{document}

\title{Magnetic monopole solutions with a massive dilaton}

\author{
         P.~Forg\'acs$^{(1)}$,
 J.~Gy\"ur\"usi$^{(2)}$ \thanks{
        corresponding author E-mail: gyurusi@poe.elte.hu}\\
       {\small $^{(1)}$ Laboratoire de Math. et Physique Theorique,} \\
       {\small CNRS UPRES-A 6083} \\
       {\small Universit\'e de Tours}\\
       {\small Parc de Grandmont, F-37200 Tours, France}\\
       {\small $^{(2)}$ Institute for Theoretical Physics,
        Roland E\"otv\"os University,} \\
       {\small  H-1088, Budapest, Puskin u. 5-7, Hungary }\\
}

\maketitle
\begin{abstract}
Static, spherically symmetric
monopole solutions of a spontaneously broken SU(2) gauge theory
coupled to a massive dilaton field are studied in detail
in function of the dilaton coupling strength and of the dilaton mass.

\end{abstract}

\vfill\eject

%\section{Introduction}
In this paper we present some results on
finite energy solutions
in an SU(2) Yang-Mills (YM) and YM-Higgs (YMH)
theory (with the Higgs field in the adjoint representation)
coupled
to a {\sl massive} dilaton field.
The present work is an extension of previous investigations with a
{\sl massless} dilaton field \cite{BizI, LavI, GYJFP1}.
In Refs.\ \cite{BizI, LavI} it has been found that the SU(2) dilaton-YM
(DYM) theory admits static, finite energy (`particle-like') solutions
(absent in the pure YM case).
They are in close analogy to the particle-like solutions found by 
Bartnik and McKinnon in an Einstein-YM (EYM) system \cite{BM}.
In Ref.\ \cite{GYJFP1} it has been shown
that in the SU(2) dilaton-YMH (DYMH) theory in addition to the analogue
of the (nonabelian) `t Hooft-Polyakov monopole \cite{THP} there is a discrete
family of finite energy solutions, which can be interpreted as radial
excitations of the monopole.
The mass scale of the radial excitations is inversely proportional
to the dilaton coupling.

In Ref.\ \cite{GYJFP1} it has also been found that there is a maximal dilaton
coupling, $\alpha_{\rm max}$, above which only an {\sl abelian} solution
exists.
Although the abelian solution (which has a simple analytical form)
is singular at the origin, its total energy is finite.
The numerical results of Ref.\ \cite{GYJFP1} indicate that the
nonabelian monopole merges with the
abelian one for a critical value of the dilaton coupling, $\alpha$.
The critical dilaton coupling, $\alpha_{\rm crit}$,
depends on the value of the Higgs self-coupling strength, $\beta$.
As found in Ref.\ \cite{GYJFP1} when $\beta$ is sufficiently small
the {\sl largest} possible value of the dilaton coupling, $\alpha_{\rm max}$,
for which
a nonabelian solution still exists is different from the {\sl critical}
value, $\alpha_{\rm crit}$, i.e.\ we have
$\alpha_{\rm max}>\alpha_{\rm crit}$.
This implies that
if $\alpha\in[\alpha_{\rm crit},\,\alpha_{\rm max}]$
a bifurcation takes place and there are two
different monopole solutions for the same value of $\alpha$. All these findings
are again in close analogy with the results of Refs.\ \cite{BFM, BFMII} found
in the EYMH case.

A massless dilaton, which necessarily appears in string theories \cite{Wi,BFQ,GSW},
violates the equivalence principle, and therefore
its (dimensionful) coupling strength
is expected to be extremely weak, of the order
of $1/M_{\rm Pl}$ where $M_{\rm Pl}$ is the Planck mass.
It is very natural
to assume, however, that the dilaton gets a mass
(possibly related to supersymmetry breaking) then, however,
there is no strong experimental constraint on the dilaton coupling.
Therefore it might be of some importance to study the effect of {\sl
mass} of the dilaton in DYMH theories. Regular and black hole solutions 
in EYM theories coupled to a massive dilaton and axion have investigated
in Ref.\ \cite{Oneil}.

We have carried out a rather thorough numerical investigation and we have
found that even if the dilaton is massive,
most phenoma associated with the presence of the dilaton
found in the massless case persist
(the existence of radial excitations, 
$\alpha_{\rm max}\ne\alpha_{\rm crit}$).
Also we have good numerical evidence that $\alpha_{\rm crit}$ 
(where the solution become abelian) is independent of the dilaton mass.
We have also investigated in detail the limit $m\to\infty$ of 
the mass of the dilaton and we were able to show its expected decoupling.  
Our numerical results show that $\alpha_{\rm max}$ grows as $m$ increases,
consistently with the expectation that 
for $m\to\infty$ $\alpha_{\rm max}(m)\to\infty$.

We have also found that not just a single 
maximal dilaton coupling, $\alpha_{\rm max}$ exists, but there are
several local extrema of $\alpha$ too, e.g.\
$\alpha_{{\rm max}_1}>\alpha_{{\rm max}_2}>\alpha_{\rm crit}>
\alpha_{{\rm min}_1}$.
These bifurcation points are more clearly distinguishable as the dilaton
mass becomes larger. This implies that there are certain values of the dilaton
coupling, $\alpha$,
where three or even more different monopole solutions exists
for the same value of $\alpha$ with different masses.

By using the minimal spherically symmetric and static ansatz
\be {A-ans}
 A_0^a=0\,,\quad A_i^a=\epsilon_{aik}{1-W(r)\over r}{x^k\over r}\,,
   \quad \Phi^a=H(r){x^a\over r}\,,\quad \varphi=\varphi(r)\,,
\ee
where $A_\mu^a$ is the gauge potential, $\Phi^a$ is the Higgs triplet
and $\varphi$ is the dilaton field,
the reduced energy functional reads:
\be {MDYMH-action-ans}
 \eqalign{
  E(\alpha,\beta^2,m^2)=\int\limits_0^\infty dr&\left[
    {1\over2}r^2\varphi'^2+{1\over2}m^2 r^2\varphi^2
    +\e^{2\alpha\varphi}\left(W'^2+{(W^2-1)^2\over2 r^2}\right)
  \right.\cr&\left.
  \quad
    +{H'^2 r^2\over2}+H^2W^2
    +{\beta^2\over8}\e^{-2\alpha\varphi}r^2(H^2-v^2)^2
  \right]\,.
 }
\ee
Varying the energy functional (\ref{MDYMH-action-ans}) we obtain the 
field equations:
\ba {MDYMH-eqs}
 (r^2\varphi')'&&=
  2\alpha\e^{2\alpha\varphi}
       \left(W'^2+{(W^2-1)^2\over2 r^2}\right)
     -{\alpha\beta^2\over4}\e^{-2\alpha\varphi}r^2(H^2-v^2)^2
     +m^2r^2\varphi\\
  \left(\e^{2\alpha\varphi}W'\right)'&&=
  W\left[
  \e^{2\alpha\varphi}{W^2-1\over r^2}+H^2
  \right]\\
  \left(r^2H'\right)'&&=
  2H\left(W^2+{\beta^2\over4}\e^{-2\alpha\varphi}r^2(H^2-v^2)\right)\,.
\ea
 By introducing the following variables:
\be {invvars}
 \eqalign{
  R=r\e^{-\alpha\varphi}:=r\phi,\quad \nu:=\alpha\varphi,\quad
  \dot{r}={dr\over d\tau}={1\over\phi}\,,
  }
\ee
 the field equations (\ref{MDYMH-eqs}) can be written as:
\ba {MDYMH-inv-eq}
 \dot R&&=1-R\dot\nu\,,\\
 {1\over\alpha^2}\left( (R^2\dot\nu)\,\dot{}+(R^2\dot\nu)\dot\nu \right)&&=
 2\left(\dot W^2+{(W^2-1)^2\over2 R^2}\right)
 -{\beta^2\over4}R^2(H^2-v^2)^2
 +{m^2\over\alpha^2}R^2\nu \e^{2\nu}\,,\\
 \ddot W+\dot W\dot\nu&&=W\left({W^2-1\over R^2}+H^2\right)\,,\\
 (R^2\dot H)\,\dot{}+R^2\dot H\dot\nu&&=
 H\left(2W^2+{\beta^2\over2} R^2(H^2-v^2)\right)\,.
\ea
 The field equations (\ref{MDYMH-inv-eq}) written in this form 
 are distinguished by the fact that for $m=0$ they contain only variables which
are {\sl invariant} under the dilatational symmetry
\be {dilsym}
r\longrightarrow \e^{\alpha c}r,\quad \varphi\longrightarrow \varphi+c\,,
\ee
which is only broken by the presence of the mass term.

Solutions regular at $r=0$ (or equivalently at $\tau=0$) must satisfy the 
following boundary conditions:
\be{boundary}
  \eqalign{
  H&=a\tau+O(\tau^3),\qquad\quad\ \  
  W=1-b\tau^2+O(\tau^4)\,,
  \cr
  R&=\tau-{\nu_1\over3}\tau^3+O(\tau^5)\,,\qquad\!
  \nu=\nu_0+{\nu_1\over2}\tau^2+O(\tau^4)\,,
   \cr
  {\rm with}\qquad
  \nu_1&={\alpha^2}\left(4b^2-{\beta^2 v^4\over12}\right)
        +{m^2\over3}\nu_0\e^{2\nu_0}\,,
  }
\ee
 i.e.\ solutions with a regular origin depend on
 three free parameters ($a$, $b$, $\nu_0$).
 The asymptotic behaviour ($r\to\infty$) of regular solutions is
almost identical to those of the $m=0$ case (Eqs.\ (10) in
Ref.\ \cite{GYJFP1}),
i.e.\ 
 for $r\to\infty$ the  b.c.\ corresponding to regular solutions are
 \begin{subeqnarray}\label{dymhinfreg}
 \varphi(r)&=&-{\alpha\over m^2r^4}+O({1\over r^6})\,,\quad m\ne0\,,\\
  %\varphi&=&\varphi_\infty-{d\over r}+O({1\over r^2})\,,\\
  W(r)&=&B e^{-r}\left(1+O({1\over r})\right)\,,\\
  H(r)&=&1-{C\over r}e^{-\beta r}\left(1+O({1\over r})
       \right)\quad{\rm for}\;\;\beta<2\\
  H(r)&=&1-{2B^2\over {(\beta^2-4)r^2}}e^{-2r}
       \left(1+O({1\over r})\right)\quad{\rm for}\;\;\beta>2\,,
 \end{subeqnarray}
 where ($B$, $C$) are free parameters and we have set $v=1$. 
We note that
$\varphi(r\to\infty)$ does not tend to zero as e$^{-mr}$,
which is the expected asymptotic behaviour of a massive field.
This `pathological' behaviour and the missing free parameter
are due to the fact that in Eq.\ 
(\ref{MDYMH-eqs}a) as $W\to0$ the $O(1/r^2)$ term dominates.
(The asymptotic behaviour of the Higgs field for $\beta>2$ in 
Eq.\ (\ref{dymhinfreg}d) is determined by the nonlinear terms 
in a similar way.) 
 
 With the use of the field eqs.\ (\ref{MDYMH-eqs}) and the boundary
 conditions (\ref{dymhinfreg}) one can easily show
 that the energy of the solution, $E(\alpha,\beta^2,m^2)$, is a 
 monotonically increasing function
 of $\beta^2$ and $m^2$ and a monotonically decreasing function of
 $\alpha$.

 In the DYM limit of the theory, i.e. 
 ($v\to0$ and $H\to0$) there are no monopoles,
 but our numerical results clearly indicate the existence of globally
 regular solutions with zero magnetic charge. The existence of these
 solutions is not surprising, as
 they are simply the generalizations of those found (also numerically)
 in the massless  case ($m=0$) in Refs.\ \cite{BizI,LavI}.
 Our results for the $m=0$ case are in complete agreement with
 those of Refs.\ \cite{BizI,LavI}.
 
\begin{figure}[htb]
 \hbox to\hsize{
   \hss
   \xname{$W$, $1/\phi$}{$r/(r+1)$}{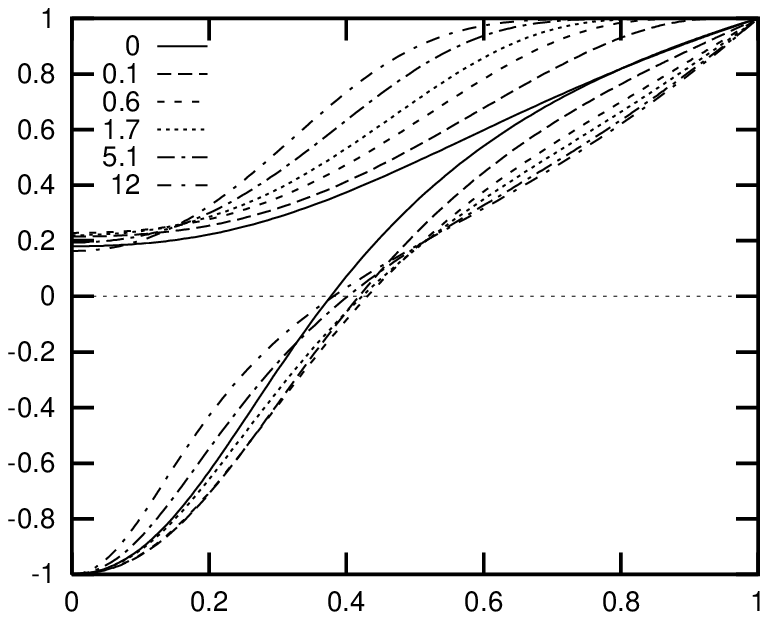}{0.5}
   \hss
   \xname{$W$, $1/\phi$}{$r/(r+1)$}{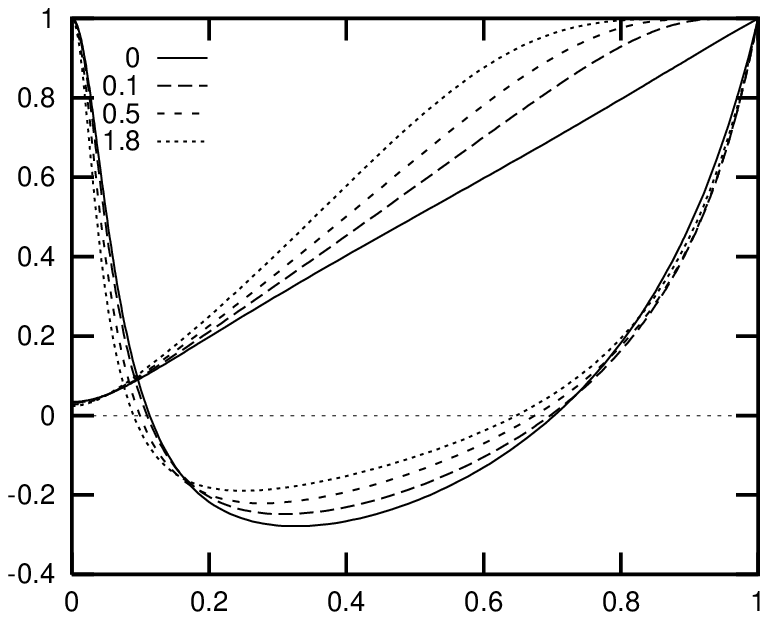}{0.5}
   \hss
 }
 \caption[DYMfig]{\label{DYMfig}
  $m^2$ dependence of the first two excited solution in the DYM limit.
 }
\end{figure}

 We have found only a slight dependence on the dilaton mass.
Although our numerical solutions 
exhibit clearly that the dilaton tends to zero more and more as 
 $m$ increases, $W$ varies very little for the considered range of values of
 $m$ ( $0\leq m\leq12$) as it can be seen on Fig.~\ref{DYMfig}.
In the $m\to\infty$ limit no nontrivial
regular solution survives, because the dilaton tends to zero everywhere
forcing $W\equiv1$. This limit is, however, singular and we expect
that for finite values of $m$ (no  matter how large) an infinity family
of regular solutions still exists.

Before considering nonabelian monopole solutions, we discuss first
the abelian case, i.e. $\varphi=\varphi_{\rm ab}$,
$W\equiv0$, $H\equiv v$ (a Dirac monopole
coupled to a massive dilaton field). 
There is a finite energy solution, with a (logarithmic) singularity
at the origin.
In the massless case $\varphi_{\rm ab}=\varphi^{(0)}$ and it is given by
\be{dil-abel}
  \varphi^{(0)}=-{1\over\alpha}\ln\abs{1+{\alpha\over r}}\,,
\ee
while for $m\ne 0$ its analytic form is not known. 
Note that for abelian solutions the only relevant parameter is 
$m\alpha$ (for $m\ne0$, $\alpha\ne0$).
It seems very likely that
the abelian solution is unique for any value of $m$. 
  
\begin{figure}[htb]
 \hbox to\hsize{
   \hss
   \xname{$1/\phi$}{$r/(r+\alpha)$}{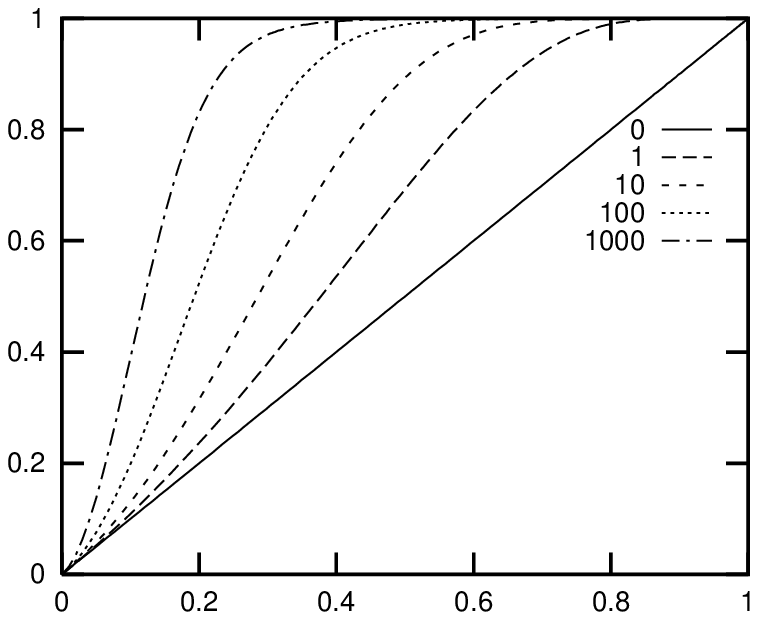}{0.5}
   \hss
   \xname{$R/(R+\alpha)$}{$r/(r+\alpha)$}{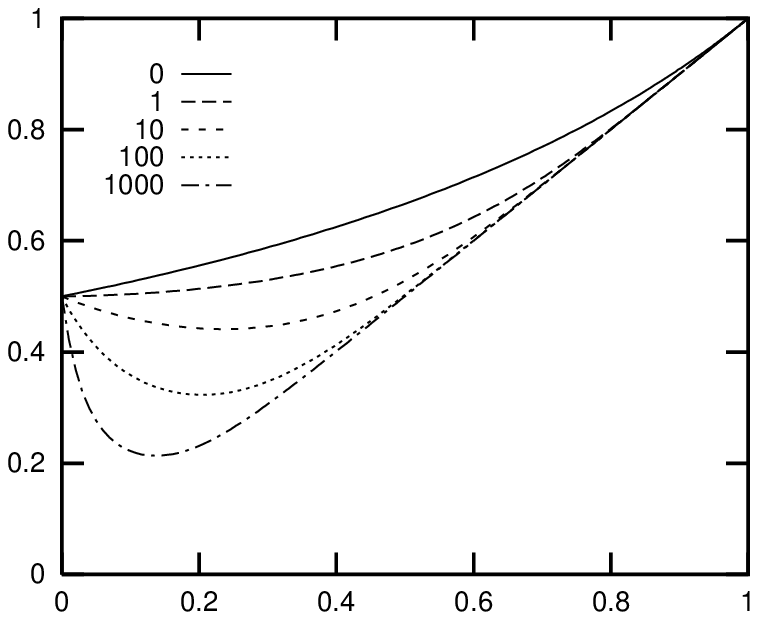}{0.5}
   \hss
 }
 \caption[abelfig]{\label{abelfig}
  $m^2$ dependence of the abelian solution for $\alpha=1$.
  See Table~\ref{abel} for their parameters. 
 } 
\end{figure}

Near the origin the power series expansion of $\varphi_{\rm ab}$ can be
written as
\be{abel-ori}
 \alpha\varphi_{\rm ab}=(1+{m^2\over4}r^2)\ln{r\over\alpha}
               +ar+({1\over2}a^2-{5\over16}m^2)r^2+\dots
\ee
 where $a$ is a free parameter. So clearly if $r\ll1/m$ the
 mass term in (\ref{abel-ori}) is negligible and the dilaton is well
 approximated by the regular solution with $m=0$, see Fig.~\ref{abelfig}.
\begin{table}[hbt]
\caption{Parameters of the abelian solution for $\alpha=1$}
\vspace{.2truecm}\label{abel}
%\noindent
\small{\hfill
\begin{tabular}{|r||r|r|}
\hline
\multicolumn{1}{|c||}{$m^2$}&
\multicolumn{1}{|c|}{$a$}&
\multicolumn{1}{|c|}{$E$}
\\ \hline\hline
    0&   1.000000& 1.000000\\
    1&   0.055585& 1.304817\\
   10&  -2.164717& 1.724591\\
  100& -10.124077& 2.594755\\
 1000& -38.701493& 4.278556\\
\hline
\end{tabular}
\hfill}
\end{table}

Next we investigate the nonabelian solutions. 
In the asymptotic region, $r\to\infty$, the behaviour of
the nonabelian solutions is determined
 by linearizing the field Eqs.\ (\ref{MDYMH-inv-eq}) 
 around the critical point 
 $\{W=0$, $H=v$, $\varphi=0\}$ for $\tau\to\infty$, 
 yielding Eqs.\ (\ref{dymhinfreg}). As in this case $\varphi\to0$
 $R\approx r$, $r\approx\tau$, $W$ and $(H-1)r$ are exponential functions of
 $\tau$.
  
 We solved the field equations (\ref{MDYMH-inv-eq}) numerically by a
 combination of a Runge-Kutta (RK) method (4th order, adaptive stepsize)
 integrating out from $\tau=0$ to some $\tau_0$ and solving by iteration
 the pertinent system of integral equations from $\tau=\infty$ to 
 $\tau_0$. 
 In order to desingularize
 the field equations (\ref{MDYMH-inv-eq}) at the origin 
 resp. at $\tau=\infty$, we have introduced the following set of
 variables:
 \{$(W-1)/R$,\, $\dot W$,\,
 $H$,\, $(rH)'$,\, $\dot\varphi$,\, $\varphi$,\,$R$,\}\ \ 
 resp.
  \{$W_\pm=W\pm\dot W$,\,$h_\pm=\beta h\pm H_+$,\,
 $\varphi_\pm=mR\varphi\pm Y$,\, $R-\tau$\}\  where $h=(H-1)R$,\, $H_+=(rH)'-1$,
 \,$Y=\varphi+R\dot\varphi$.\\
 (For $\beta=0$ we used $R^2\dot H$ instead of $h_-$, and for $m=0$ we
 replaced $\varphi_-$ by $R^2\dot\varphi$.)
 The corresponding eqs.\ for $\beta\not=0$ and $m\not=0$ are the following:
 % ($e,v=1$)
\ba{inteq}
  &&
  \eqalign{
  W_\pm=&\mp\int_{1/\tau_{\pm}}^{1/\tau}
      \left[
            W\left({W^2-1\over R^2}+{h\over R}(h/R+2)\right)
              -\alpha\dot W\dot\varphi\right]t^2\e^{\pm (\tau-1/t)} d\,t
  \cr&         +C_{W_\pm}\e^{\pm (\tau-\tau_\pm)}
  }
  \\&&
  \eqalign{
  h_\pm=&\mp\int_{1/\tau_{\pm}}^{1/\tau}
     \left[ {2W^2H\over R}+{\beta^2 h^2\over2R}(h/R+3)
            \mp\alpha\beta h\dot\varphi\right]t^2\e^{\pm\beta(\tau-1/t)} d\,t
  \cr&         +C_{h_\pm}\e^{\pm\beta(\tau-\tau_\pm)}
  }
  \\&&
  \eqalign{
  \varphi_\pm=&\mp\int_{1/\tau_{\pm}}^{1/\tau}
           \left[ {\alpha\over R}\right( 2\dot W^2+{(W^2-1)^2\over R^2}
          -{\beta^2h^2\over4}(h/R+2)^2 \left)\phantom{H\over H}
         \right.\cr&\quad\quad\left.  +m^2 K(\e^{2\alpha\varphi}-1)
        \mp m K\alpha\dot\varphi \right]t^2\e^{\pm m(\tau-1/t)} d\,t
        +C_{\varphi_\pm}\e^{\pm m(\tau-\tau_\pm)}
  }
  \\&&
  R-\tau=\int_{1/\tau_{-}}^{1/\tau}
          \alpha\, R\,\dot\varphi\, t^2 \,\,d\,t+C_{R-\tau}
\ea
 where $\tau_{+}=\infty$, $\tau_{-}=\tau_0$.
 The $C_k$s are arbitrary constants determined by the
 boundary conditions (\ref{boundary},\ref{dymhinfreg}) ensuring regularity
 at $\tau=0$ and at $\tau=\infty$.
 In order to supress the divergent modes in $\{W_+,h_+,\varphi_+\}$
 at $\tau=\infty$ one has to choose $C_{W_+},C_{h_+},C_{\varphi_+}=0$.
 The remaining four constants $\{C_{W_-},C_{h_-},C_{\varphi_-},C_{R-\tau}\}$
 are determined by 
 $C_{W_-}=W_-(\tau_0)$, $C_{h_-}=h_-(\tau_0)$,
 $C_{\varphi_-}=\varphi_-(\tau_0)$, $C_{R-\tau}=R(\tau_0)-\tau_0$,
 where $\{W_-(\tau_0), h_-(\tau_0),\varphi_-(\tau_0), R(\tau_0)-\tau_0\}$
 are given by the RK procedure, guaranteeing regularity of the functions at
 $\tau=0$. The integral eqs.\ (\ref{inteq}) are then solved by iteration
 yielding a regular solution for $\tau\in[\tau_0,\infty]$. Then the shooting
 parameters for the RK are adjusted so that the resulting values
 $\{W_+(\tau_0),h_+(\tau_0),\varphi_+(\tau_0)\}$
 match those calculated from the solution of the integral equations.
 
\begin{figure}[htb]
 \hbox to\hsize{
   \hss
   \xname{$W$, $H$}{$r/(r+1)$}{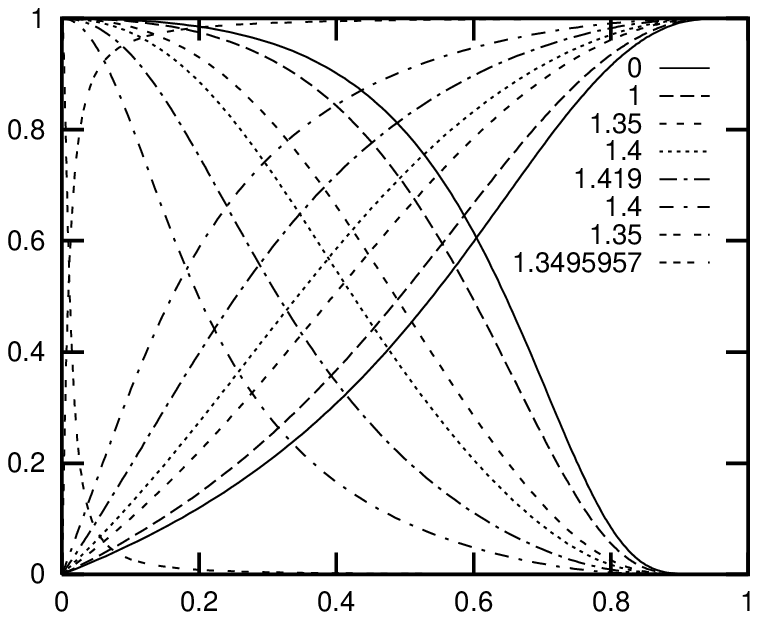}{0.5}
   \hss
   \xname{$1/\phi$}{$r/(r+1)$}{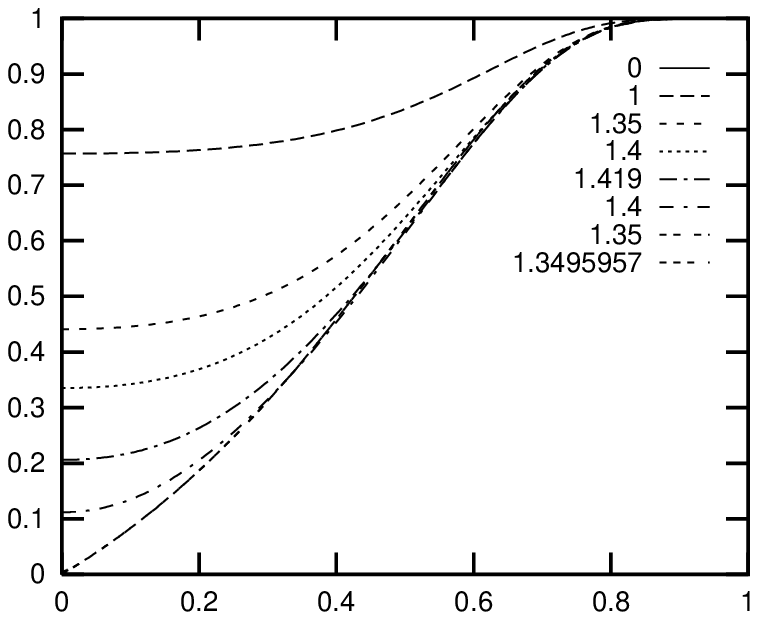}{0.5}
   \hss
 }
 \hbox to\hsize{
   \hss
   \xname{$W$, $H$}{$\bar r/(\bar r+1)$}{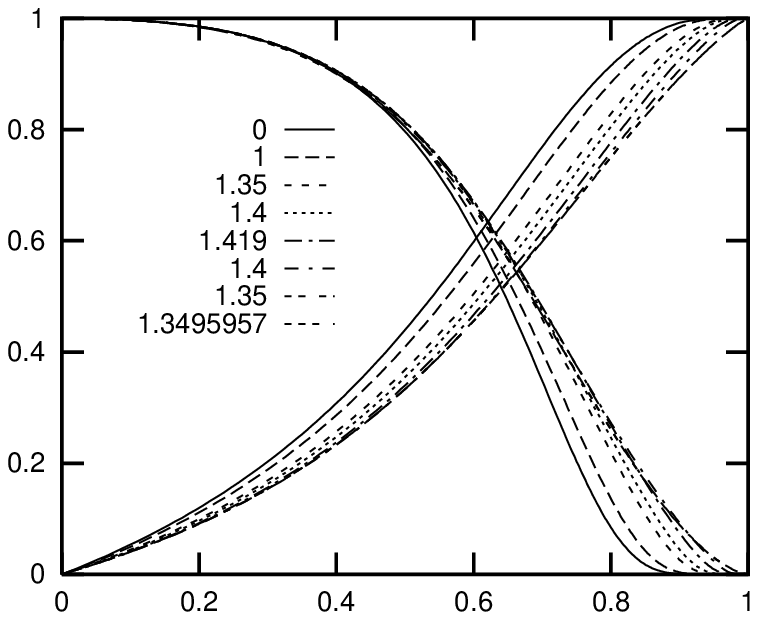}{0.5}
   \hss
   \xname{$\varphi_{\rm sh}$}{$\bar r/(\bar r+1)$}{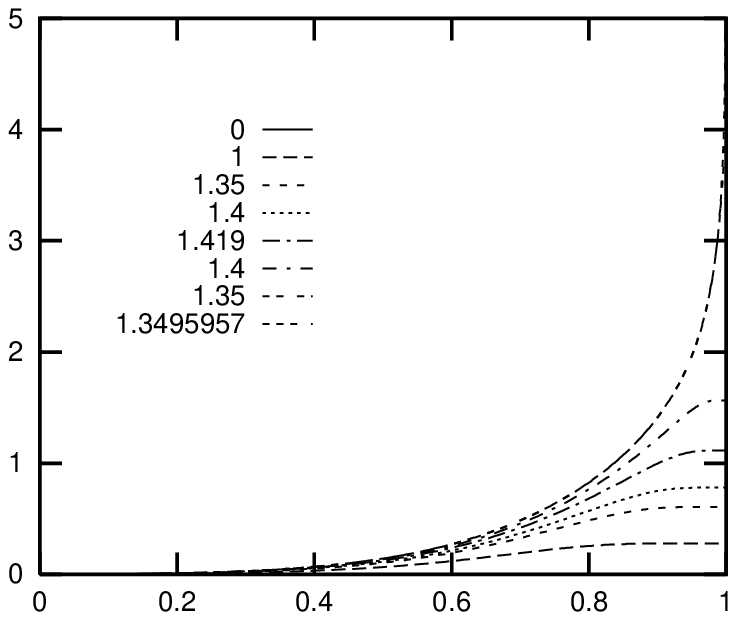}{0.5}
   \hss
 }
 \caption[fundmon1]{\label{fundmon1}
  The $\alpha$ dependence of the fundamental monopole solution for $m=1$,
  $\beta^2=0.1$. Above the exterior, below the interior solution is depicted.
 }
\end{figure}

 Let us now present our numerical results in some detail.
 First of all the fundamental nonabelian monopole solutions, which
 tend to the 't Hooft-Polyakov monopole as $\alpha\to0$
 exists only up to
 a maximal value of the dilaton coupling strength $\alpha$,
 $\alpha_{\rm max}(\beta,m)$. For $\alpha>\alpha_{\rm max}$ only the abelian
 solution \{$\varphi\equiv\varphi_{\rm ab}$, $W\equiv0$, $H\equiv1$\} seems to
 exist. The $\alpha$ dependence of the solution is, however, quite complex as
 the nonabelian solution does not necessarily cease to exist for
 $\alpha=\alpha_{\rm max}$. In fact that value of $\alpha$ 
 (which value we call critical, $\alpha_{\rm crit}$) for which the
 nonabelian solution merges with the abelian one, 
 does not necessarily coincide with $\alpha_{\rm max}$. 
 As $\alpha$ approaches its critical value, $\alpha_{\rm crit}$, the 
 dilaton field 
 becomes divergent at the origin, whereas $W(r)\to0$ and $H(r)\to1$.
  This behaviour is shown on
 Fig.~\ref{fundmon1} where the fundamental monopole is plotted as
 $\alpha$ varies from 0 to $\alpha_{\rm crit}$.
 Note that  although the solution becomes singular its total energy
 remains finite.
\begin{table}[hbt]
\caption{Parameters of the fundamental monopole solution for $m=1$,
 $\beta^2=0.1$}
\vspace{.2truecm}
\label{fund1}
%\noindent
\small{\hfill
\begin{tabular}{|l||r|r|r|r|r|}
\hline
\multicolumn{1}{|c||}{$\alpha$}&
\multicolumn{1}{|c|}{$a$}&
\multicolumn{1}{|c|}{$b$}&
\multicolumn{1}{|c|}{$-\varphi_0$}&
\multicolumn{1}{|c|}{$\alpha E$}
\\ \hline\hline
          0&  0.4821127&  0.2381420&  0.0000000& 0.0000000\\
          1&  0.4481335&  0.2370648&  0.2787202& 1.0605540\\
       1.35&  0.4021605&  0.2368443&  0.6065505& 1.3627414\\
        1.4&  0.3874162&  0.2378367&  0.7807487& 1.3967137\\
      1.419&  0.3750471&  0.2400726&  1.0195001& 1.4078919\\
      1.419&  0.3719369&  0.2411243&  1.1136470& 1.4078993\\
        1.4&  0.3650065&  0.2464302&  1.5649554& 1.3999542\\
       1.35&  0.3662630&  0.2550531&  3.8698184& 1.3849301\\
  1.3495957&  0.3662688&  0.2546880&  9.3420476& 1.3848394\\
\hline
\end{tabular}
\hfill}
\end{table}

 The critical value of $\alpha$ depends strongly on $\beta$, but it
 does not seem to depend on the mass of the dilaton, $m$.
 This  latter phenomenon can be understood as follows.
 By performing the transformation (\ref{dilsym}) we define
  $\varphi_{\rm sh}(r)=\varphi(r)-\varphi_0$ where $\varphi_0:=\varphi(0)$
  guaranteeing $\varphi_{\rm sh}(0)=0$ and then the field
 eq.\ (\ref{MDYMH-eqs}a) takes the form:
  \be{shifteq}
  (\bar r^2\varphi_{\rm sh}')'=
  2\alpha\e^{2\alpha\varphi_{\rm sh}}
       \left(W'^2+{(W^2-1)^2\over2 \bar r^2}\right)
     -{\alpha\beta^2\over4}\e^{-2\alpha\varphi_{\rm sh}}\bar r^2(H^2-1)^2
     +m^2 \e^{2\alpha\varphi_{0}}\bar r^2(\varphi_{\rm sh}+\varphi_{0})\,,
  \ee 
   where $\bar r=re^{-2\alpha\varphi_{0}}$ and the prime denotes now
 derivation with
   respect to $\bar r$. Note that $\varphi_{\rm sh}(\bar r)\to -\varphi_{0}$
   as $\bar r\to\infty$.
  The remaining field eqs.\ (\ref{MDYMH-eqs}b,\ref{MDYMH-eqs}c)
  are form invariant.
   
  Using the shifted field, $\varphi_{\rm sh}$, in order to obtain
  a globally regular solution with
  some {\sl fixed} asymptotic value of $\varphi_{\rm sh}$, -$\varphi_{0}$,
  one also has to vary $\alpha$ in addition to the shooting parameters 
  $a$, $b$. (One still has to supress three divergent modes).
  Then the critical value of $\alpha$ is defined by taking 
  the limit $\varphi_{0}\to-\infty$.
  Then in Eq.\ (\ref{shifteq}) for any, {\sl fixed} value of $\bar r$ 
  the terms proportional to $m^2$ tend to zero as $\varphi_{0}\to-\infty$.
  From this it follows that $\alpha_{\rm crit}$
  is determined by the {\sl massless} equations. Note that it is essential to
   keep $\bar r$ fixed when taking the $\varphi_{0}\to-\infty$ limit.
 In this limit one obtains nontrivial
  $\{\bar W(\bar r)$, $\bar H(\bar r)\}$ which are globally regular 
  and a logarithmically
  divergent $\bar \varphi(\bar r)$ as $\bar r\to\infty$. Since for 
  any {\sl fixed} value of $\bar r$ $r\to0$ as $\varphi_{0}\to-\infty$
  this solution will be referred to as the `interior' solution.
  It
  corresponds to a singular solution of Eqs.\ (\ref{MDYMH-eqs}),
  $W(r)\equiv0$, $H(r)\equiv1$ for $r\ne0$
  with a logarithmically diverging dilaton field at $r=0$ which we call
 `exterior' solution. Note that while
  the interior solution is nonabelian, the exterior one is, however, abelian.

\begin{table}[hbt]
\caption{$m^2$ and $\beta^2$ dependence of $\alpha_{\rm max}$,
$\alpha_{\rm min}$, $\alpha_{\rm c}$}
\vspace{.2truecm}
\label{alpha-m}
%\noindent
\scriptsize{\hfill
\begin{tabular}{l|l||l|l|l|l|}
                  &$m^2$&$\beta^2$=0& 0.07      & 0.1        & 1          \\
\hline\hline
                  & 0   & 1.408800 &1.361925    &     ---    &1.206183    \\
                  & 0.1 & 1.411778 &1.361912    &     ---    &1.206184    \\
$\alpha_{\rm max}$& 0.5 & 1.471277 &1.374598    &1.359389    &1.209233    \\
                  & 1   & 1.586124 &1.439274    &1.419335    &1.244148    \\
                  & 5   & 2.314522 &1.965262    &1.924131    &1.612840    \\
\hline\hline
                  & 0   &      --- &1.361900    &     ---    &    ---     \\
                  & 0.1 &      --- &1.361898    &     ---    &    ---     \\
$\alpha_{\rm min}$& 0.5 &      --- &1.361843    &1.349560    &    ---     \\
                  & 1   &1.399088  &1.361764    &1.349553    &1.20618034  \\
                  & 5   &1.39581   &1.36137     &1.34947     &1.20617980  \\
\hline\hline
                  & 0   &      --- &    ---     &     ---    &   ---      \\
                  & 0.1 &      --- &    ---     &     ---    &   ---      \\
$\alpha_{\rm max2}$& 0.5&      --- &    ---     &     ---    &   ---      \\
                  & 1   &      --- &1.361902156 &1.349595705 &   ---      \\
                  & 5   &      --- &1.361902253 &1.349596618 &   ---      \\
\hline\hline
                  & 0   & 1.3993823&1.361902    &1.349595679 &1.20618036  \\
                  & 0.1 & 1.3993823&1.361902    &1.349595679 &1.20618036  \\
$\alpha_{\rm c}$  & 0.5 & 1.3993823&1.361902    &1.349595679 &1.20618036  \\
                  & 1   & 1.3993823&1.361902154 &1.349595679 &1.20618036  \\
                  & 5   & 1.3993823&1.361902154 &1.349595679 &1.20618036  \\
\hline
\end{tabular}
\hfill}
\end{table}

 With our numerical procedure we found that in fact 
 there is a local minimum of $\alpha$, $\alpha_{\rm min}$,
 and that even a second local maximum of $\alpha$ exists.
 We remark that some values (for $m=0$) differ from our previous results 
 as given in Table~2 of ref.\ \cite{GYJFP1}.
 The present data is more precise due to the use of
 the integral equations around infinity.

 The numerical data show that the local minimum of
$\alpha$ exists
 for a wide range of $\beta$, especially in the case of large dilaton masses.
 This local minimum exists even in the $m=0$ case, but
for larger $m$ the
 difference between the different extrema of $\alpha$ 
 and of $\alpha_{\rm crit}$ is more visible. This is
 the reason why $\alpha_{\rm min}$  has not been found in Ref.\
  \cite{GYJFP1},
  where the present matching procedure has not been used. 
  It is quite possible
 that even more local extrema of $\alpha$ exist but 
 our numerical data is not precise enough to see them. 
 To investigate this question a better method is needed as
  one has to approach $\alpha_{\rm crit}$ very closely where
 our numerical errors become too large.  
 As $\beta$ increases (with $m$ fixed) $\alpha_{\rm min}$ and
 $\alpha_{\rm max_2}$ gets closer.
 The mass and the $\beta$ dependence of the different extrema of $\alpha$
 together with those of $\alpha_{\rm crit}$ are given in Table~\ref{alpha-m}.
 We also notice that $\alpha_{\rm max}$ grows
 quite rapidly as $m$ increases while the critical alpha remains unchanged.
\begin{figure}[hbt]
 \hbox to\hsize{
   \hss
   \xname{$a$}{$\alpha$}{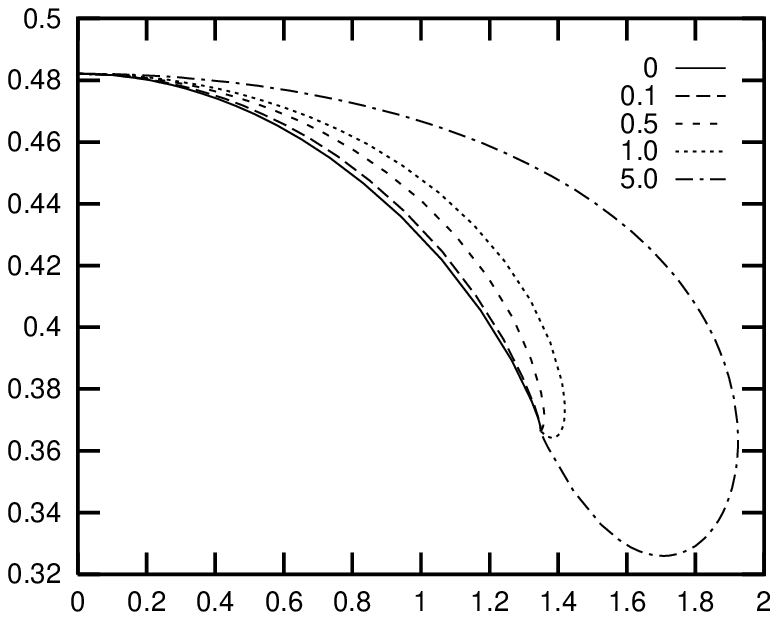}{0.5}
   \hss
   \xname{$b$}{$\alpha$}{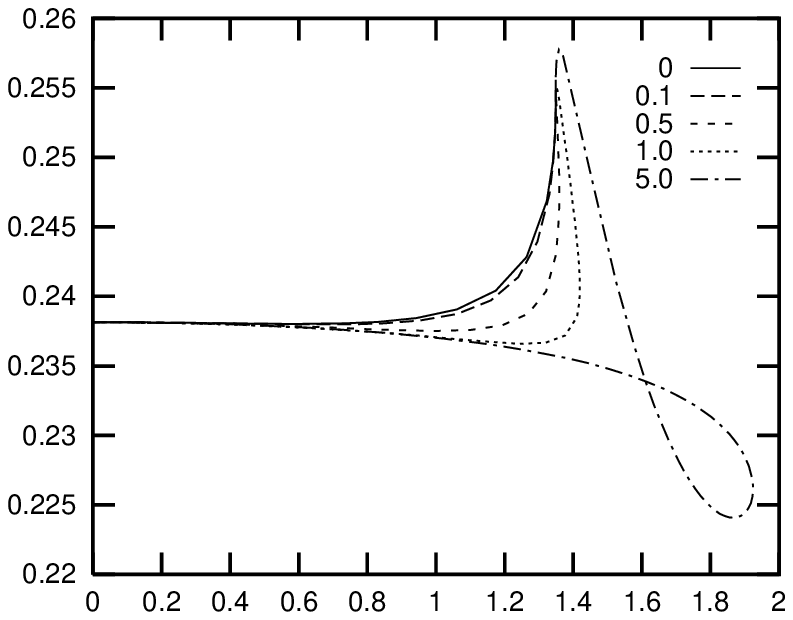}{0.5}
   \hss
  }
 \hbox to\hsize{
   \hss
   \xname{$1/\phi_0$}{$\alpha$}{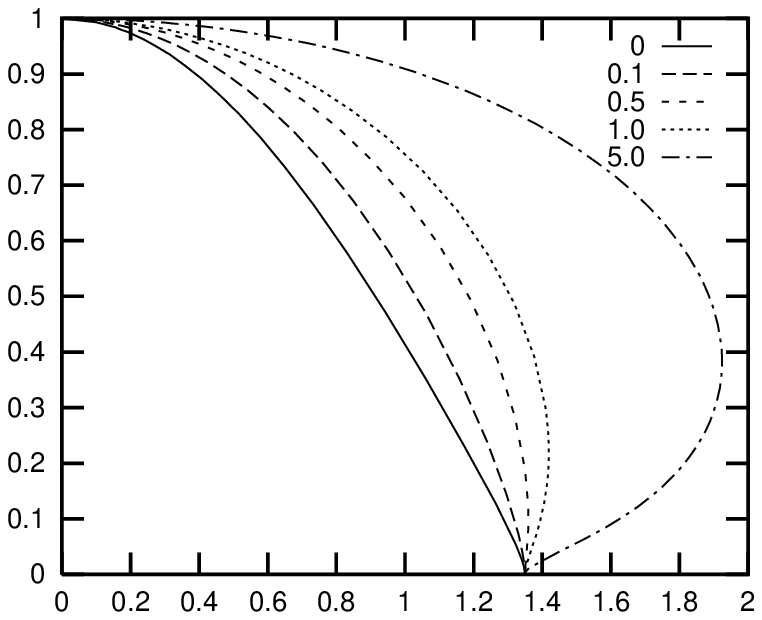}{0.5}
   \hss
   \xname{$E$}{$\alpha$}{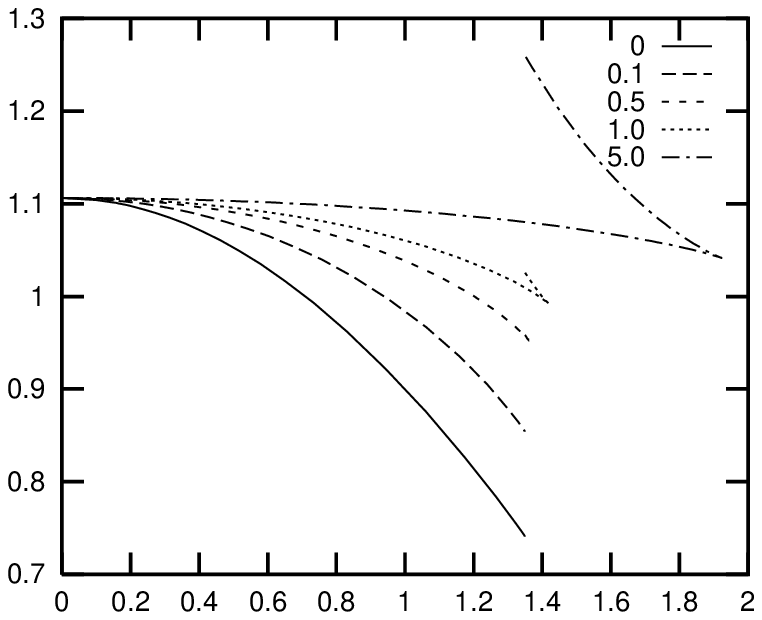}{0.5}
   \hss
 }

 \caption[fund01ab]{\label{fund01ab}
  The $\alpha$ dependence of the shooting parameters $a$, $b$, $1/\phi_0$
 and the energy of the
  fundamental monopole with $\beta^2=0.1$ for some $m^2$.
 }
\end{figure}
 This behaviour is illustrated on Fig.~\ref{fund01ab}
 where the $\alpha$ dependence of the shooting parameters 
 $a$, $b$, $e^{\alpha\varphi_0}$ and the energy
 of the solution, E, for different values of $m$ are plotted. 
 It is clearly visible that they all
 become more and more constant for 
 $ \alpha\ll\alpha_{\rm max}$ as $m$ grows.  
 So the solution becomes more and more independent of $\alpha$ for
 $\alpha\in[0\,,\alpha_{\rm max}]$ as $m$ increases and gets
 closer and closer to the 't Hooft-Polyakov monopole. 
 The numerical results indicate that in the limit $m\to\infty$ 
 the functions $W$ and $H$
approach those corresponding to the 't~Hooft-Polyakov monopole 
 while $\varphi\to0$. 

\begin{figure}[htb]
 \hbox to\hsize{
   \hss
   \xname{$W$, $H$}{$r/(r+1)$}{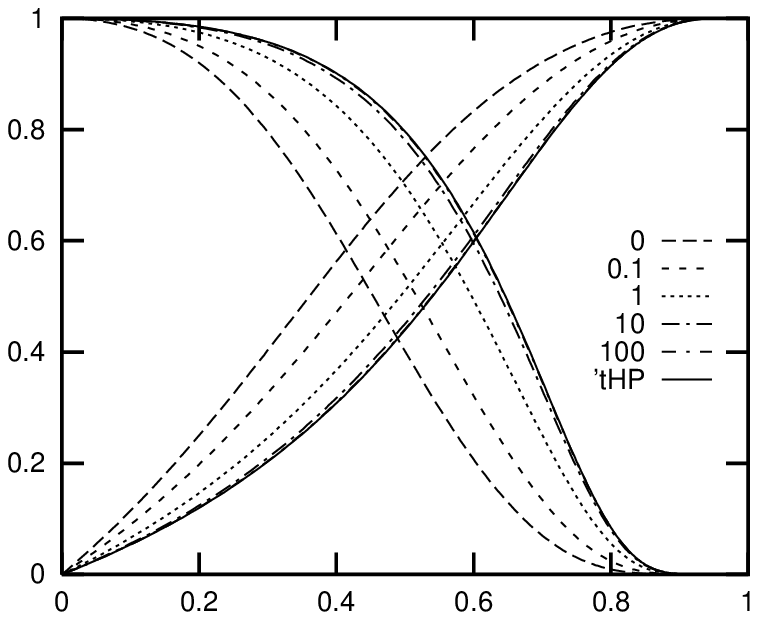}{0.5}
   \hss
   \xname{$\tilde\varphi$}{$r/(r+1)$}{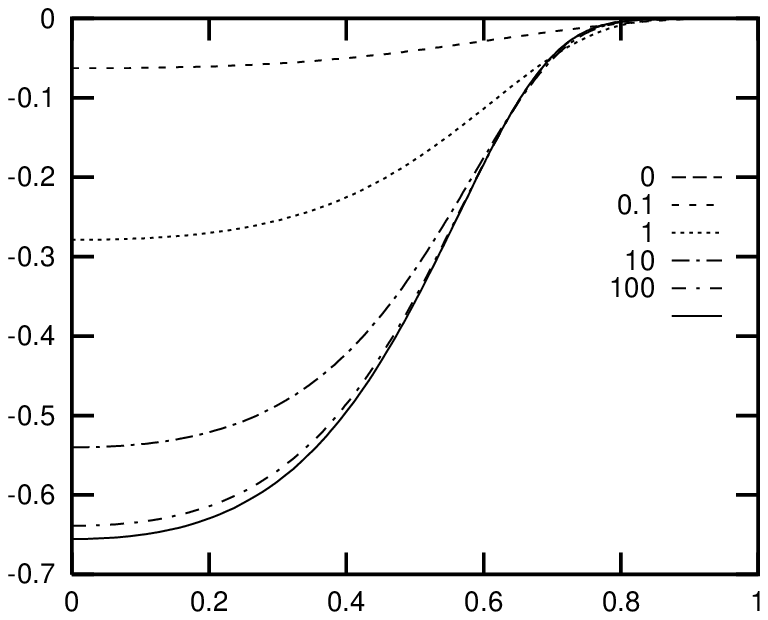}{0.5}
   \put(-41,92){\makebox(0,0)[r]{$\tilde\varphi_{\rm inf}$}}
   \hss
 }
 \caption[fund0]{\label{fund0}
  The $m^2$ depedence of the fundamental monopole solution for $\alpha=1$,
  $\beta^2=0.1$. $\tilde\varphi=m^2\varphi/\alpha$
  and 'tHP denotes the 't Hooft-Polyakov monopole solution.
  The corresponding parameters are given in Table~\ref{fund-m}.
 }
\end{figure}

To understand the $m\to\infty$ limit somewhat better 
let us define $\tilde\varphi(r)$ by $\varphi=\alpha\tilde\varphi/m^2$. 
 Assuming now that $\varphi(r)\ll1$ and keeping only the leading terms in $m^2$
 it follows from the field equations 
 (\ref{MDYMH-eqs}) that $W\approx W_{\rm thp}$, $H\approx H_{\rm thp}$ where
 $W_{\rm thp}$, $H_{\rm thp}$ denote the 't Hooft-Polyakov solution and 
 $\tilde\varphi\approx\tilde\varphi_{\rm inf}$ where
 $$
 \tilde\varphi_{\rm inf}=-{1\over r^2}\left(2W_{\rm thp}'^2
 +(W_{\rm thp}^2-1)^2/r^2-
 {\beta^2\over 4}r^2(H_{\rm thp}^2-1)^2\right)\,.
 $$
The numerical results show that already for
$m^2=10$\,\,\, $\tilde\varphi_{\rm inf}$ approximates $\tilde\varphi$ very well,
see Fig.~\ref{fund0}. The
shooting parameters also tend to the appropriate ('t~Hooft-Polyakov) values.
Fig.\ \ref{fund01ab} also exhibits the tendency that
the shooting parameters and the energy of the solution depends less and less on
$\alpha$ as $m$ increases.

So the above result supports the expectation that in the limit $m\to\infty$
the dilaton decouples while tending to zero and
$W$, $H$ converging to the 't~Hooft-Polyakov monopole solution.
Note that the growing of the maximal $\alpha$, $\alpha_{\rm max}$,
as $m$ becomes larger is also consistent with the expected decoupling.

\begin{table}[hbt]
\caption{Parameters of the fundamental monopole solution for $\alpha=1$,
 $\beta^2=0.1$, together with the corresponding values for the
 't Hooft-Polyakov monopole.
 The starred value stands for the limiting
 value of $-\tilde\varphi(0)$  for $m^2\to\infty$ i.e.\ for
 $-\tilde\varphi_{\rm inf}(0)=12b^2-\beta^2/4$\,.}
\vspace{.2truecm}
\label{fund-m}
%\noindent
\small{\hfill
\begin{tabular}{|r||r|r|l|r|}
\hline
\multicolumn{1}{|c||}{$m^2$}&
\multicolumn{1}{|c|}{$a$}&
\multicolumn{1}{|c|}{$b$}&
\multicolumn{1}{|c|}{$-\varphi_0$}&
\multicolumn{1}{|c|}{$E$}
\\ \hline\hline
 0  & 0.4292657&  0.2386866&  0.8816368&  0.9001767\\
 0.1& 0.4321814&  0.2384289&  0.6263968&  0.9845152\\
 1  & 0.4481335&  0.2370648&  0.2787202&  1.0605536\\
 10 & 0.4726173&  0.2373279&  0.0539949&  1.0988876\\
 100& 0.4808578&  0.2380086&  0.0063885&  1.1054006\\
'tHP& 0.4821127&  0.2381420& $0.6555393^*$&  1.1061955\\
\hline
\end{tabular}
\hfill}
\end{table}

\begin{figure}[htb]
 \hbox to\hsize{
   \hss
   \xname{$W$, $H$}{$\ln(r)$}{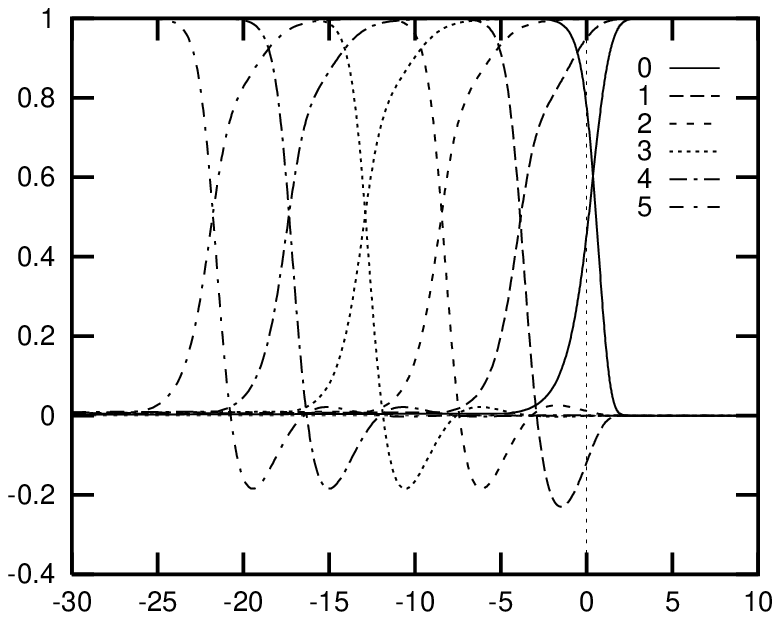}{0.5}
   \hss
   \xname{$\varphi$}{$\ln(r)$}{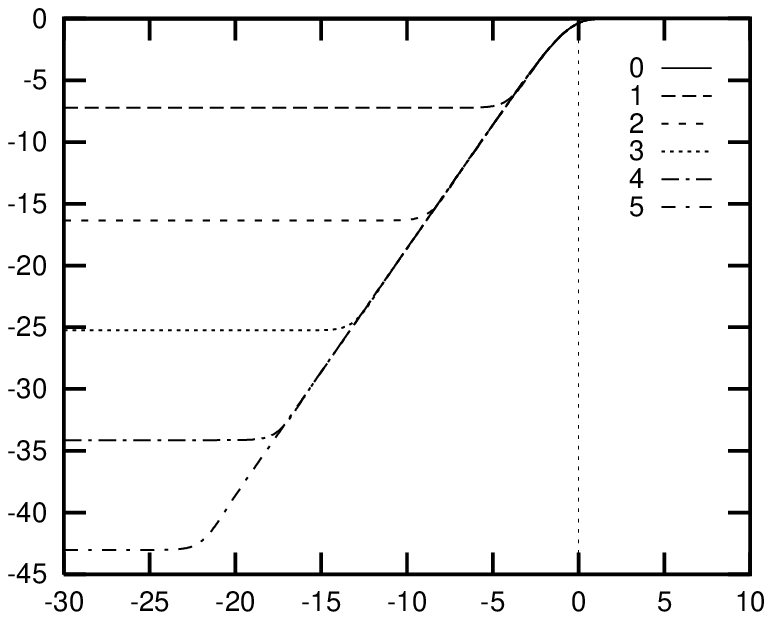}{0.5}
   \hss
 }
 \caption[excit]{\label{excit}
  The first five excited monopole solution for $\alpha=0.5$,
  $\beta^2=0.1$, $m^2=1$. Their parameters are given in Table~\ref{family}.
 }
\end{figure}

In addition to the fundamental monopole there is a
 discrete family of nonabelian solutions 
 characterized by the number of zeros of $W$, $n$.
 They can be interpreted
 as radial excitations of the fundamental ($n=0$) monopole.
 See Fig.~\ref{excit}. 
\begin{table}[hbt]
\caption{The parameters of the fundamental and first five excited monopole
solutions for $\alpha=0.5$, $\beta^2=0.1$, $m=1$}
\vspace{.2truecm}
\label{family}
%\noindent
\small{\hfill
\begin{tabular}{|r||r|r|r|r|}
\hline
\multicolumn{1}{|c||}{$n$}&
\multicolumn{1}{|c|}{$a$}&
\multicolumn{1}{|c|}{$b$}&
\multicolumn{1}{|c|}{$-\varphi_0$}&
\multicolumn{1}{|c|}{$\alpha E$}
\\ \hline\hline
0& 0.4746849&  0.2378870&   0.1120060&  0.5478783\\
1& 0.8255632&  1.5068349&   7.2301508&  1.1638519\\
2& 0.8669493&  1.5525918&  16.3566790&  1.1741097\\
3& 0.8676030&  1.5530864&  25.2510967&  1.1742216\\
4& 0.8676121&  1.5530923&  34.1370513&  1.1742229\\
5& 0.8676122&  1.5530924&  43.0228028&  1.1742229\\
\hline
\end{tabular}
\hfill}
\end{table}
 One can understand these oscillating solutions by 
 linearizing the field eqs.\ (\ref{MDYMH-inv-eq}) around 
  $\{W=0$, $H=v$, $\varphi=1/\alpha\ln(r/\alpha)\}$
 i.e.\ ($R\approx\alpha$) in the region $r\ll1/m$ and $r\ll\alpha$
 corresponding to $\tau\approx\alpha\ln r+$const.
 (For $m=0$ $\{W=0$, $H=v$, $R=\alpha$,
 $\dot\nu=1/\alpha$\}  is a critical point of Eqs.\ (\ref{MDYMH-inv-eq}).)
  The linearization yields the following behaviour:
 $\{W(\tau)\propto{\rm e}^{\lambda_{W}\tau}$, 
 $(H(\tau)-1)\tau\propto{\rm e}^{\lambda_{H}\tau}$, 
 $\varphi(\tau)\propto{\rm e}^{\lambda_{\varphi}\tau}$\}; 
 where the exponents, $\lambda_i$, are given as
\be{lambda-ori}
 \eqalign{
  \lambda_W&=1/\alpha(-1/2\pm\sqrt{\alpha^2-3/4}),
  \cr
  \lambda_H&=1/\alpha(1/2\pm\sqrt{\alpha^2\beta^2+1/4}),\quad
  \lambda_\varphi=1/\alpha(1,-2)\,.
  }
\ee
The amplitude of the oscillations of $W$ tends to zero as $1/\sqrt r$ and 
 their frequency is independent from the mass of the dilaton.
 From Eqs.\ (\ref{lambda-ori})
  one expects that solutions with an oscillating
  $W$ and a logarithmically growing  dilaton  
 (in the region where $r\ll$min$\{\alpha,1/m\}$) exist only for
$\alpha^2<3/4$ . For larger values of $r$ 
 the behaviour of $\varphi$ changes 
 from $\varphi\propto\ln(r)$ to $\varphi\propto1/r^4$
 and $W$ stops to oscillate.
 We expect that globally regular solutions with an arbitrary number of
 oscillations exist for $\alpha^2<3/4$.
 
 In the massless ($m=0$) case we have found excited solutions
 (with $n\ne0$) only up to $\alpha=\sqrt3/2$ independently of $\beta$ and $n$. 
 But with an increasing mass
 we have also observed solutions with $n=1$ for $\alpha>\sqrt3/2$ 
 (E.g.\ $\alpha=0.98$, $m^2=10$ for $\beta^2=0.1$).
 In fact there is also a maximal and a critical value of 
 $\alpha$ for the excitations similarly to the $n=0$ case. 
 Our numerical
 data is, however, not precise enough to allow for a good 
 determination of $\alpha_{\rm crit}$, (for $n\ne0$)
 but they are consistent with $\alpha_{\rm crit}=\sqrt3/2$ independently of
 $m$ and $n$.
 This is in agreement with the argument given for the mass independence of
 $\alpha_{\rm crit}$.
 We remark that the maximal value of $\alpha$ for solutions with $n=1$
 has always been less than one. 
 The energy of the excitations increases with $n$ and it
 is always larger than that of the fundamental monopole.
 The first zero of an oscillating solution  moves closer and closer 
 to $r=0$ as the number of oscillations increases. 
 The excited solutions depend only slightly on the dilaton mass,
 similarly to what has been seen in the massive DYM theory ($v\to0$ limit). 
 
 We expect that the excited solutions are unstable against spherical
 perturbations, because of the established instability 
 of their counterparts in the massless DYM theory \cite{BizI,LavI}. 
 We also expect the fundamental monopole solutions 
 to become unstable after the bifurcation point at $\alpha=\alpha_{\rm max}$,
 analogously to the case of gravitating monopoles \cite{H}.
\vskip1truecm
\noindent
{\bf Acknowledgements}
\vskip.4truecm
 We would like to thank D. Maison and P. Breitenlohner for helpful discussions.

\end{document}